\documentclass[twocolumn]{aa}

\usepackage{txfonts}

\usepackage{graphicx}

\usepackage[authoryear]{natbib}

\begin{document}
   \title{The shortest period M dwarf eclipsing system BW3~V38, II: determination
   of absolute elements  \thanks{based on observations collected at the European Southern 
			 Observatory, La Silla, Chile}
            }

   %\subtitle{}

   \author{C. Maceroni
          \inst{1}
          \and
          J. Montalb\'{a}n \inst{2}
          }

   \offprints{C. Maceroni}

   \institute{INAF - Osservatorio Astronomico di Roma
              via Frascati 33, I-00040 Monteporzio C. (RM) - Italy\\
              \email{maceroni@mporzio.astro.it}
        \and
            Istitut d' Astrophysique et G\'{e}ophysique Universit\'{e} de Li\`{e}ge,
			All\'{e}e du 6 A\^{out}, B-4000 Li\`{e}ge, Belgium\\
            \email{j.montalban@ulg.ac.be}
             }

   \date{Received March 30, 2004; accepted June 23, 2004}

   \abstract{The spectroscopic data for the short-period ($0 \fd 1984$)
eclipsing binary V38, discovered by the OGLE micro-lensing team in Baade's Window 
field BW3, are analyzed. Radial velocity curves are derived from mid-resolution
spectra obtained with EMMI-NTT at ESO - La Silla,  and  a simultaneous solution of the
existing light curve by OGLE and of the new radial velocity curves is obtained.
The system is formed by almost twin M3e dwarf components that are very close, 
but not yet in contact. The spectra of both dwarfs show signatures of the presence
of strong chromospheres. Spectroscopy definitely confirms, therefore, what was suggested
on the basis of  photometry: BW3~V38 is indeed a unique system, as no other 
similar binary with M components and in such a tight orbit is known. 

Within the limits posed by the relatively large errors,  due to the combined effect of 
system faintness and of the constraints on exposure time,
the  derived physical parameters seem to agree with the relations obtained from the other few 
known  eclipsing binaries with late type components  (which indicate a discrepancy between
the available evolutionary models and the data  at $\sim$ 10\% level). A possible 
explanation is the presence of strong magnetic fields and  fast rotation (that applies
to the BW3~V38 case as well).
A simple computation of the system secular evolution by angular momentum loss
and spin--orbit synchronization shows that the evolution of a system with M dwarfs
components is  rather slow, and indicates as well a possible reason why  
systems similar to BW3~V38 are so rare.
   }
\titlerunning{Absolute elements of BW3~V38}
   \maketitle
%
%________________________________________________________________

\section{Introduction}
\object{OGLE BW03~V038} ( BW3~V38, for brevity, through this paper; 
$\alpha _{\mathrm{2000}} = 18^{\mathrm{h}}04^\mathrm{m}44 \fs 19$, $\delta _{\mathrm{2000}} = -30\degr 09\arcmin 05\arcsec$,
$P = 0 \fd 19839$, $\mathrm{I_{max}} = 15.83$,
$\mathrm{(V-I)_{max}}=2.45$, $\Delta \mathrm{I} = 0.78$)  
appears in the second instalment of the OGLE Catalog of periodically
variable stars \citep{uda2} and is therein
classified, on the basis of  light curve shape, as a contact
binary of  W~UMa-type (EW).  It  was later rejected, however,  
from the bona-fide sample of contact systems in the OGLE fields, 
which was selected by  \citet{ruc97} by means of an objective Fourier-based classification of
the light curves.  The system was found
marginally outside the  region that, in the  space of parameters defining the light
curve properties, corresponds to the contact configuration.

\citet[][ hereafter paper I]{MR97} analyzed the OGLE photometric data, providing
a  solution of the I-band light curve and studying in  detail the
related uncertainties. They reported, as well, that the  system is undetected 
in the ROSAT survey, with an upper limit to the X-ray flux of $1.8 \times 10^{-13}$ erg 
cm$^{\mathrm{-2}}$ s$^{\mathrm{-1}}$ (this
result was  not surprising on the basis of the system properties and of a 
rough estimate  of its distance of $\sim 400$ pc). 

 The de-reddened V--I color of BW3~V38 ($\mathrm{(V-I) _0}=2.3-2.4$, paper I),
its light curve solution and  the short period  suggest a configuration with two similar 
very close and fast rotating M-dwarfs. 
This makes of BW3~V38 an extremely  interesting object on several grounds.

While a few  eclipsing binaries with M-type components are known at present 
(see the recent paper by \citet{rib03} for a summary) BW3~V38 is the only one with so  tight
an orbit,  and this, according to the current knowledge on close binary formation,
is very unlikely to be its original configuration, as ``in situ" formation is ruled out
for such small separations (see Section 4). 
 
 We think, therefore, that BW3~V38 started its MS life with a larger separation and has evolved 
through the combined effect of angular momentum loss  by magnetic braking (AML) 
and of spin--orbit coupling. The presence of a strong magnetic field was suggested indeed 
by the typical magnetic activity signatures found  in the photometric data, as
 the solution of the OGLE light curve (paper I) 
required a large dark spot on the stellar surface of the primary component. 
We expected, therefore, to detect other typical activity indicators in the spectra. 
 
   In synchronized binaries the net effects of AML   are shrinkage
of the orbit, orbital period decrease and faster spinning of the components. The
secular evolution of the orbital
period is  strictly related to the efficiency of  magnetic braking.

 At the time when the available data were rather sparse and inhomogeneous, the  
lack of  close binaries with late (M) type components could be explained in terms 
of a selection effect against the discovery of intrinsically faint objects. However, 
the modern large surveys for
variability (such as OGLE) have shown that the effect is real, as  these systems 
are indeed very rare, even in large,  homogeneous and at some level complete datasets.

 The dearth of such systems implies then  some evolutionary effect. Either, for some 
configuration, the orbital evolution should become so fast to prevent system detection or,
on the contrary, it should be very slow.   
 The latter hypothesis seems more likely in light of the current understanding
of the AML processes in late-type single stars. 

Studies of cluster  \citep[][ and references therein]{Teal00}
and of field M dwarfs  \citep{delal98} of different ages favor the hypothesis of a spin-down
time-scale that increases with decreasing mass (and show a saturation of the activity
indicators with rotation).   This evidence has been related to the
properties of the  $\alpha\Omega$ dynamo that operates at the  interface between the
radiative core and the convective envelope, and that is expected to work up to spectral type M3 
(or maybe later for magnetic dwarfs, see \citet{MMD01}). The dynamo and the braking efficiency
increase -- at fixed radius -- with  faster rotation, until  a "saturation" of the mean field is reached  
at  rotational velocities that slightly depend on the sample,  but are of the order of 
10 km~s$^{\mathrm {-1}}$.  
Looking instead at the behavior with  spectral type, as the angular momentum
is essentially  extracted from the convective envelope, which deepens with
advancing spectral type, later stars  have longer spin-down timescales.
Finally, in fully convective stars a further reduction of AML is expected, 
as  the dynamo changes in nature (perhaps the small-scale turbulent dynamo of \citet{ddr93}).

In synchronized binaries with late type components, that -- at variance with single
stars -- rotate faster with increasing age, both effects contribute to lengthen the 
"contact" timescale. 
The studies of the period distribution of various samples of close binaries with solar-type 
components  confirmed the abovementioned scenario, suggesting as well a saturated braking 
law \citep{R83, MV91, Ste95, MR99}.

The expected physical properties of BW3~V38, as guessed in paper I solely on the basis of
the  photometry, put the system exactly at the proposed border of the efficient braking 
and make the system of paramount importance for the study  of magnetic braking mechanisms
and the effects of fast rotation in late-type dwarfs.
  Its location in the C-M diagram is also of great interest, as our system lies
exactly midway between the two  -- and for a long
time the only -- other known eclipsing binaries with late dwarf components, 
\object{YY Gem} and \object{CM Dra}, 
fundamental lower--MS calibrators,  and next  to 
its recently discovered, and presumably younger ``twin", \object{CU Cnc} \citep{delal99}.

Therefore, a spectroscopic study of this unique system was  absolutely needed.
In this paper we report the results of the observing runs aimed to get mid-resolution
spectra of the binary. Section 2 describes the details of data acquisition and
reduction, Section 3 presents the simultaneous solution of the light and radial
velocity curves and a discussion of the system physical properties, Section 4 gives
an outline of  the possible orbital evolution, followed by  
the conclusions.        
%__________________________________________________________________

\section{The spectroscopic observations}

BW3~V38 is a rather difficult target for radial velocity determination. 
Its  period requires short exposures to avoid Doppler smearing of spectral lines.
Assuming, as a rule of thumb, that an acceptable exposure should last no more than 
5\% of the orbital period (i.e. 0.05 in phase), the maximum duration shall not
exceed 900$^\mathrm{s}$. Therefore, because of its faintness, a medium-large class telescope
is required. Furthermore, the field  is relatively crowded and the binary has a nearby 
 (at $\sim 2 \arcsec$ distance) and brighter (by $\sim 2$ mag ) 
optical companion.
 
 BW3~V38 was observed at ESO -- La Silla during two half nights, in August 2001 
(a previously scheduled run in July 2000 was completely lost because of bad weather).  
A total of 17 long slit spectra were obtained, at various orbital phases, with the ESO  
3.5m New Technology Telescope (NTT) and the ESO Multi--Mode Instrument (EMMI) 
in the red mid-resolution spectrographic mode. The slit width was fixed at $1 \arcsec$.  
The spectra are  640 \AA\ wide and are centered around the H$\alpha$ line 
($\lambda=6563$ \AA\,). The selected grating (\#6)  has a nominal resolution
of 5500 (at 6000 \AA\ and with the $1 \arcsec$-wide slit) and a dispersion of 12.5 
\AA\ mm$^{-1}$ (or $0.3$ \AA\ pix$^{-1}$). 
 \begin{figure}
    \resizebox{\hsize}{!}{\includegraphics{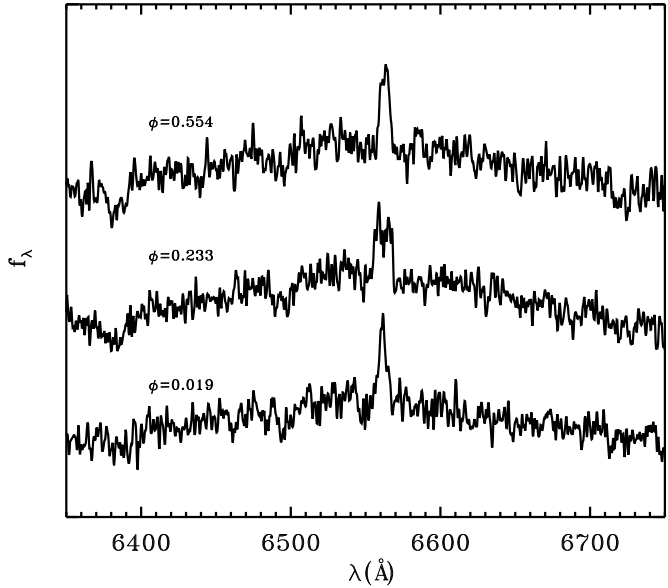}}
     \caption{Three smoothed spectra of BW3~V38, from bottom to top: around primary eclipse,
	 quadrature and secondary eclipse. The spectra are
	  normalized to the highest continuum level. The splitting of 
	  H$\alpha$ at quadrature indicates the presence 
      of  strong chromosphere in both components.
	  }
         \label{spectra}
   \end{figure}
   \begin{figure}
	\resizebox{\hsize}{!}{\includegraphics{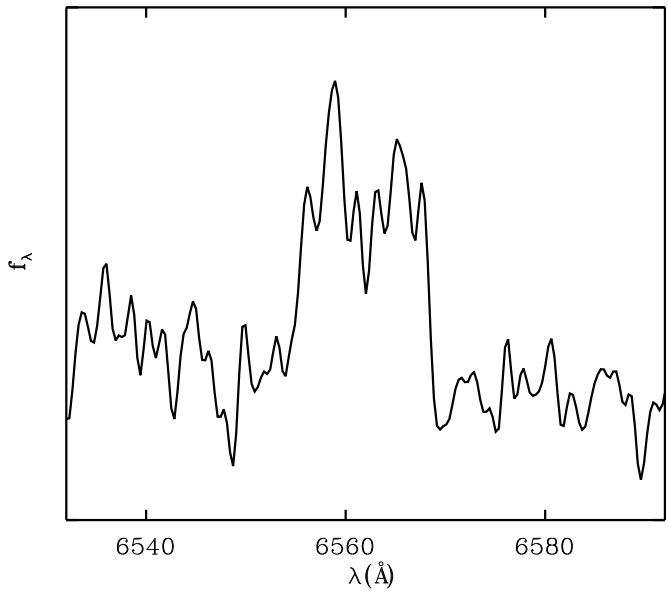}}
      \caption{Detail of the H$\alpha$ region in the spectrum at phase $\phi=0.233$, note
	  the partial reversal of the profiles. 
	  }
         \label{spdet}
   \end{figure}
 The spectra of a few other K--M dwarfs, to be used as templates, were acquired as well; 
the stars were chosen from the lists of \citet{kirk91} and \citet{kirk94}. 
Finally, we selected the two templates that looked more similar to the spectra of the
binary and of the optical companion, i.e. \object{GJ~729} (M3.5~V) and \object{GJ~775}
(K5~V), for which accurate radial velocities  were recently obtained by
\citet{Nidal02}. 

Because of the above-mentioned constraints the exposures were limited to 
900$^\mathrm{s}$ and the slit  was aligned along the direction connecting 
our main target with the optical companion  (and not --  as usual -- according to 
the parallactic angle). In this  way the spectra of both  objects could 
simultaneously be acquired and the contribution of the close-by
star could be  taken  into account in data reduction, providing -- as an extra bonus --   
a useful internal check.
  
 The  data reduction was performed by means of standard
techniques (the long slit reduction routines of the IRAF \footnote{IRAF is 
distributed by the National Optical Astronomical Observatories, which
are operated by the Association of the Universities for Research in 
Astronomy, Inc., under cooperative agreement with the National Science Foundation} 
package).
The steps were bias subtraction, flat fielding, (partial) cosmic ray
removal,  one-dimensional spectrum extraction  and wavelength calibration.   
The radial velocities were obtained by  standard cross correlation technique
(as provided by the IDL 5.5 and IRAF packages) after a few preliminary steps of:
noise reduction (smoothing by convolution with a 3-pix FWHM gaussian curve, 
i.e. smaller than the slit projection), normalization 
to a pseudo-continuum  (obtained by means of the  IRAF continuum polynomial fitting 
and rejection algorithms) and transformation to the heliocentric system.
The orbital phase determination showed an excellent agreement with the ephemeris 
published by \citet{uda2}; in fact the same binary was briefly observed
during the second phase of the OGLE experiment, a few years later, 
and  no measurable period change or phase shift was found (Udalski, private communication). 

At this stage, an inspection of the radial velocities of the optical companion
revealed the existence of a non-negligible nightly positive trend in the derived 
velocities (with a range of $\simeq 30$ km~s$^{-1}$), corresponding to $\sim 2$ pixels. 
While its origin remained unknown, but could presumably be ascribed to mechanical 
problems, the trend could  be removed completely by a careful use of the night sky 
emission lines. 
 The dispersion solution was  recomputed, using for wavelength reference the night-sky 
lines, which could be identified using the spectral atlas of OH and O$_2$ emission 
lines by \citet{ost96}. Each spectrum was wavelength-calibrated with its own
sky lines  (with the exclusion of the bright template stars, whose exposure were too 
short to get a usable sky level; in that case the  nearest exposure of the binary was 
used). In this way the trend completely disappeared  and the standard deviation of the 
radial velocity of the optical companion  decreased from 9.3 to  2.2 km~s$^{-1}$.

Figure \ref{spectra} shows three spectra of the variable close to the two minima and to
quadrature.
The typical S/N ratio of the binary spectra is  $\sim 10$, i.e. still acceptable 
for radial velocity determination.  However, the task was not expected to be easy, because of 
the late spectral type and the related problems of continuum normalization, and because of the 
high rotational velocity of the components.

On the basis of Figure \ref{spectra} and of the parameters we know from the light curve solution 
of paper I (in particular the high inclination that assures almost total eclipses), we
can conclude  that the binary is indeed formed by two active similar late-type dwarfs. 
While we do not have in our  spectral range  features that allow a quantitative 
spectral classification (as, for the red part of a late dwarf spectrum,  the TiO5 index of \citet{ghr95}),
the general shape of the spectrum and the visibility of the TiO bands at 6569 and 6651 \AA\ 
suggest a M3 V classification. 

The second firm conclusion we can draw from the figure is that 
both components are dMe stars,   as a splitting of  H$\alpha$ in emission  is clearly visible
at quadrature.
   \begin{figure}
\centerline{\resizebox{8.8cm}{!}{\rotatebox{0}{\includegraphics{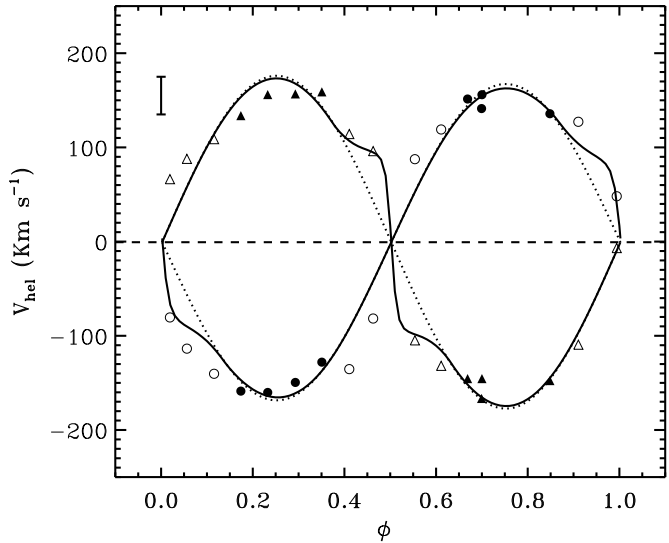}}}}
      \caption{The radial velocity curve of BW3~V38. Dots indicate the primary and triangles
	  the secondary velocity curve, the full line the result of the simultaneous solution,
	  the dotted line the stellar barycentric radial velocities. Empty symbols correspond to 
	  points that were given half
	  weight in the simultaneous solution. The vertical bar is the mean uncertainty of the
	  observed points.}
         \label{rv}
   \end{figure}
    \begin{figure}
	\resizebox{\hsize}{!}{\includegraphics{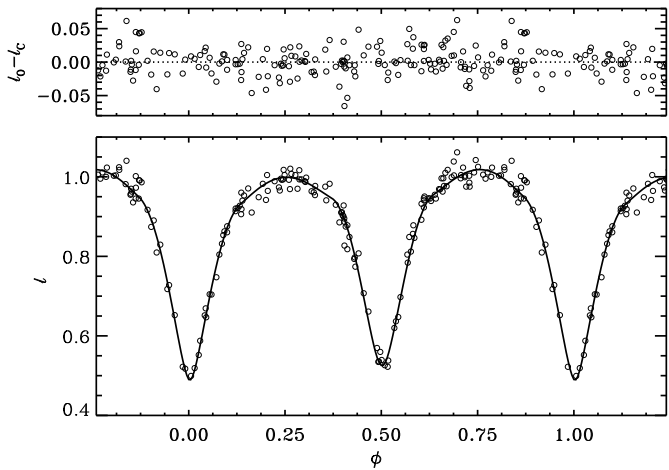}}
      \caption{The  new light curve solution BW3~V38 (lower panel) and the deviations. Paper I
	  solution is virtually indistinguishable from the new one and was not drawn, to avoid
	  confusion.}
         \label{lc}
   \end{figure}
   \begin{figure}
	\resizebox{\hsize}{!}{\includegraphics{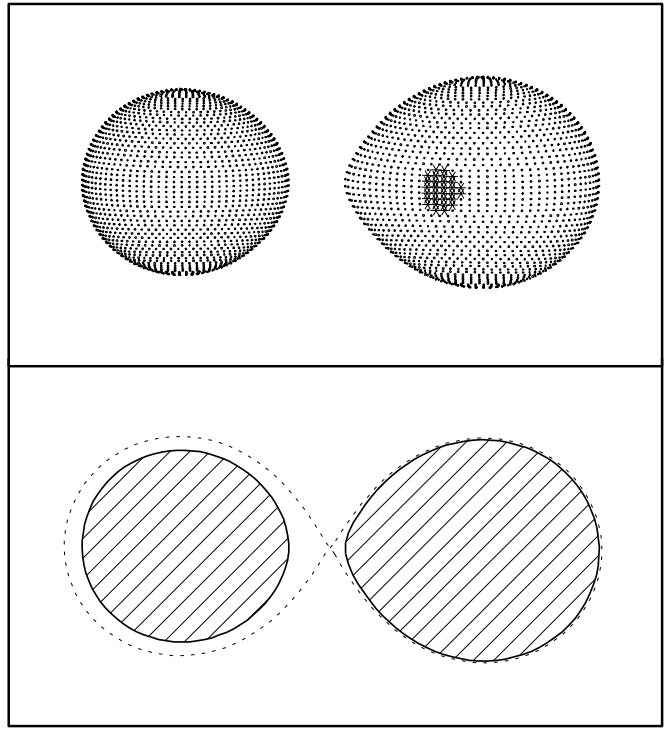}}
      \caption{ The 3-D drawing at orbital phase 0.25 
	  (upper panel) and the equatorial section (lower panel) of BW3~V38. The dashed line
       in the lower panel indicates the critical inner Roche lobe.}
	  \label{conf}
   \end{figure}
For both components the  H$\alpha$ equivalent width around quadrature is $~\simeq 2.7$~\AA\,, 
a value indicating the presence of strong chromospheres. 
Among dwarfs with H$\alpha$ in emission, however,  higher EW values are found, especially in 
components of wider (and presumably younger) binary systems \citep{ghr02}, typical values 
span the range  0-10 \AA\,. 

The age of BW3~V38 is unknown, as well 
as its population characteristics; however its location in the galactic plane (paper I) and the low value of at least 
the barycentric velocity  (see  Table~\ref{sol}) offer a weak support for a young/\,intermediate disk population. 
Regardless, our system must be  significantly older than the very similar but wider system 
\object{CU Cnc} (\object{GJ~2069A}). This too is formed  by two M3 dwarfs, but has an orbital period of 
$2 \fd 77$ and an estimated age, from the likely association with the Castor moving group,
of only $\simeq 300$ My \citep[][ hereafter R03]{rib03}. According to \citet{ghr02}, the 
H$\alpha$ EW of \object{CU Cnc} is $\cong$ 4.6 \AA\, and its activity level, as estimated from the ratio 
$L_{\mathrm{H \alpha}}/L_{\mathrm{bol}}$,  is very high. R03 reports individual EW values for each component
(of 3.85 and 4.05 \AA\ for the primary and the secondary) and a value of 
$L_x/L_{\mathrm{bol}}\cong 10^{-3}$, that presumably marks the activity saturation limit
\citep{delal98}.

The  shape of the H$\alpha$ profile of BW3~V38, that around quadrature
clearly shows a partial central reversal for both components (Figure \ref{spdet}),  
indicates a lower activity  level than  typically observed in most close binaries 
containing dMe dwarfs, which usually show a peaked profile  \citep{stha86}.
  
Figure \ref{rv} shows the radial velocity curves, as derived by the cross correlation technique.
The cross correlation functions were fitted with gaussian curves, leaving as free parameters
the position and height. The amplitudes were estimated from the fit   close to minimum
phases (the light curve solution indicates almost total eclipses of two similar size 
components) and kept constant for the other orbital phases. 
As the least square fitting routine
provided, as usual,  unrealistically small errors, a   value between 20 and 30 km~s$^{-1}$ 
(depending on the component and phase) was estimated by means of the IRAF implementation of 
the  \citet{TD79} algorithm, where the velocity errors are also computed, based on the fitted 
peak height and the antisymmetric noise.  
 This estimate tells us that the accuracy of the derived orbital elements will not be high enough to 
provide a -- badly needed -- improvement of the lower main 
sequence calibration.
However, a full exploitation of the existing data will  improve our knowledge of this unique 
binary system. 

  \section{Physical properties of BW3~V38}

A simultaneous solution of the radial velocity curves and the OGLE I--band light curve
was computed by means of the Wilson and Devinney code \citep[][ in its 2001 
distribution \footnote{the code is kindly made available by the author at
http://www.astro.ufl.edu/pub/wilson}]{WD71}.
The input parameters for the light curve solution were taken from paper I and 
those for the radial
velocity curves were obtained by a preliminary sine fit of the velocity curves. 
Looking for further information on the component color and temperature, we checked if the 
system was detected in the 2MASS survey \citep{Skal95}, but the only source at the appropriate 
location  corresponds  to a K-type star, i.e. to the brighter optical companion 
(unfortunately its optical separation with the binary is comparable to the survey true PSF 
resolution of $\simeq 2.5 \arcsec $ (FWHM)). We kept therefore the primary effective temperature
derived in paper~I. 

   \begin{table}
      \caption[]{Simultaneous light and radial velocity curve solution of BW3~V38}
         \label{sol}
     $$ 
\begin{array}{p{0.2\linewidth}ll}
%\begin{array}{lll}
\hline
\noalign{\smallskip}
Parameter &    {\mathrm{this\: paper}} & \mathrm{paper\: I} \\
\noalign{\smallskip}
\hline
\noalign{\smallskip}
i (degr)		& 85.51  \pm	0.84  &	85.7   \pm 1.2	\\
q          		& 0.950  \pm	0.067 & 0.77   \pm 0.09 \\
$\Omega _1 $ 		& 3.71   \pm	0.11 &	 3.48  \pm 0.13	\\
$\Omega _2 $  		& 4.02   \pm	0.19  & 3.55   \pm 0.25 \\
T$_2$ (T$_1$=3500)& 3448 	 \pm	11	  & 3459   \pm 15 	\\  
L$_1$/(L$_1$+L$_2$)	& 0.594   \pm	0.031 &  0.597 \pm 0.036\\
$\Delta \phi $		& 0.0023 \pm 0.0005 & 0.0020 \pm 0.0006\\

\noalign{\smallskip}
\hline
\noalign{\smallskip}
r$_1$ (side) & 0.372 \pm  0.008 & 0.380 \pm 0.011 \\
r$_2$ (side) & 0.323 \pm  0.028 & 0.322 \pm 0.050  \\
\noalign{\smallskip}
\hline
\noalign{\smallskip}
A (R$_\odot$) &1.355  \pm 0.066 & \\
K$_1$ (km~s$^{-1})$   & 167   \pm 8     & \\
$\gamma$ (km~s$^{-1})$& -0.6  \pm 5.9 & \\
            \noalign{\smallskip}
            \hline
         \end{array}
     $$ 
   \end{table}
   \begin{figure*}
\centering
	\includegraphics[width=17cm]{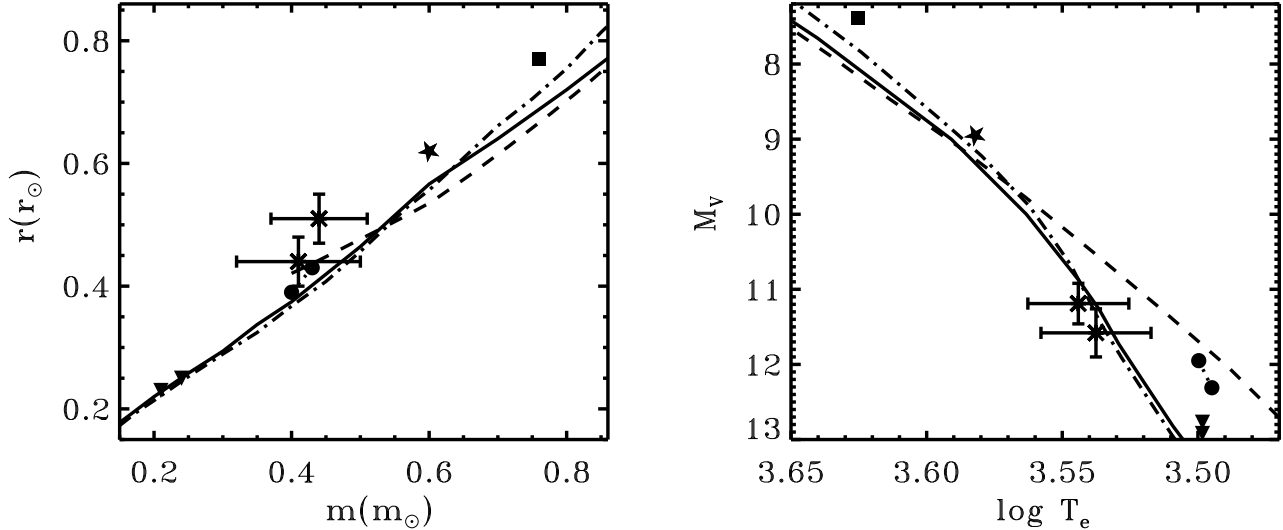}
      \caption{ The location of BW3~V38 components (points with error bars) in the mass radius 
	  and in the $\log T_{\mathrm{e}}$ -- $\mathrm{M_V}$ diagrams. The locations of the other known 
	  eclipsing binaries with M-dwarf component(s) are also shown: the filled dots indicate
	  CU Cnc components (R03), the starred dot the secondary star of V818 Tau, the
	  square YY Gem mean conponent (TR02), the triangles CM Dra components  \citep{Lacy77,metal96}.  
	  The theoretical isochrones (1 Gy, solar chemical composition) are from 
	  \citet{Dant} (continuous line), \citet{baral98} (dashed-dotted line) and \citet{Yal01} 
	  (dashed line).
	  }
	  \label{theo}
   \end{figure*}
  
As in paper I, the photometric observations were given equal weights, while the radial 
velocities were unequally weighted ($w_i=0.5$ for phases close to the photometric minima,
see Fig. \ref{rv}).
The curve dependent weights (required to weigh the different sets
of observations with respect to each other)
were assigned, respectively, on the basis of the standard deviation of the preliminary sine fit for the radial
velocity curves and of the local fit at  maxima for the light curve. 
The corresponding values are $\sigma _{\mathrm{rv}}=25$ km~s$^{-1}$
and $\sigma _{\mathrm{lc}}=0.02$ (in light units). The iterative solution 
used the multiple subset  scheme \citep{WB76}. 
The adjusted parameters were the orbital inclination, i, the mass ratio\footnote{ 
Star ``1" is defined in the code as the component eclipsed at primary minimum,  in our case $q<1$ 
and  star ``1", the primary, is as well the more  massive component.}, $q=m_2/m_1$, the surface reduced potentials,
$\Omega _{\mathrm{1,2}}$, the secondary effective temperature,$\mathrm{T_2}$,
the primary fractional luminosity, $\mathrm{L_1/(L_1+L_2)}$, 
the orbit semi-axis, A, the barycentric velocity $\gamma$, and the phase shift $\Delta \phi$.
The final solution is shown Figures 
(\ref{rv}, \ref{lc}) and the final parameter values are collected in Table \ref{sol}.  The latter contains
as well  the component ``side" fractional radii (i.e. the fractional radii in the orbital plane and
at 90$\degr$ 
from the axis connecting the star barycenters) and the primary radial velocity curve amplitude,
$\mathrm{K_1}$. The other non-adjusted parameters defining the solution maintained  the same values 
of paper I, namely the linear limb darkening coefficients $x_{\mathrm{1,2}}=0.56$ at $\lambda=7520$ \AA\,   
(and $x_{\mathrm{1,2}}=0.65$ at H$\alpha$), the gravity darkening coefficients $\beta _{\mathrm{1,2}}=0.32$, the albedos
$a_{\mathrm{1,2}}=0.5$.
Moreover, the dark spot of paper-I  was kept unchanged in the simultaneous solution and is 
defined by longitude and angular radius, respectively of 290 and 11 degrees, and the 
temperature factor with respect to the unspotted surface, $T_s/T_*=0.8$.
All the errors reported in the tables are standard errors.

      \begin{table}
      \caption[]{Derived physical quantities for BW3~V38 components.}
         \label{physq}
     $$ 
\begin{array}{p{0.2\linewidth}cc}
\hline
\noalign{\smallskip}
Parameter (solar units) &   {\mathrm{primary}} & {\mathrm{secondary}} \\

\noalign{\smallskip}
\hline
\noalign{\smallskip}
m   			  & 0.44 \pm  0.07  &  0.41  \pm 0.09  \\
R 		 		  & 0.51 \pm  0.04  &  0.44  \pm 0.06  \\  		
M$_{\mathrm{bol}}^{\mathrm{a}}$& 8.39 \pm  0.08  &  8.78  \pm 0.13  \\
$\log$ g          & 4.66 \pm  0.13  &  4.77  \pm 0.21  \\

\noalign{\smallskip}
\hline
\noalign{\smallskip}
%\end{tabular}
         \end{array}
     $$ 
\begin{list}{}{}
\item[$^{\mathrm{a)}}$] based on M$_{\mathrm{bol},\odot}=4.75$
\end{list}
   \end{table} 

The introduction of the radial velocity curves in the solution slightly changes the system 
configuration (Fig. \ref{conf}) with respect to that obtained in paper I. 
The most affected parameter 
is -- as one could expect -- the mass ratio, that increases from $q=0.77$ to $q=0.95$.
The other adjusted parameters vary as predicted by the correlation plots of 
paper~I,  that were computed by means of bootstrap "resampling". 
The configuration is still a detached one, but the primary is slightly closer to the critical lobe
(the ratio between the mean stellar radius and that of the critical surface increases from 
$r/r_R=0.95$ to $r/r_R=0.98$), while the secondary is somewhat more inside  ( $r/r_R$
decreases from 0.90 to 0.86). The  increase in relative radius is balanced by the decrease
of  secondary temperature, so that the fractional luminosities are almost unchanged.

It is well known that in detached and  partially eclipsing systems the light curve shape is 
only weakly dependent on the mass ratio and that there is a rather strong correlation between
the geometric parameters.  This explains why  the maximum deviation between  the new and the 
paper I synthetic light curves, corresponding to
the two solutions, is of only  0.005 mag  (and the mean difference of just 0.001 mag). 

The derived absolute dimensions are presented in Table \ref{physq}. The uncertainties 
on the physical parameters are unfortunately large, nevertheless   
the location of the components in the 
H~--R and in the mass radius diagrams (Figure \ref{theo}) are consistent with those of typical 
M-dwarfs. 
 
Figure \ref{theo}  shows a comparison between the location  of the known double-lined 
eclipsing binaries with M dwarf components and three  theoretical isochrones of different
authors (\citet{Dant,baral98,Yal01}).

The  binary systems  are: the already mentioned \object{CU Cnc (R03)},  \object{CM Dra}
 \citep{Lacy77,metal96} \object{YY Gem}  and \object{V818 Tau} \citep{TR02}, whose  
 secondary component is an M dwarf. 
It has to be stressed that CM Dra is  a non-homogeneous system with respect to the rest of the 
sample (and to the theoretical isochrones), as it seems to be a population II object 
\citep{Lacy77}.  The physical parameters of the various objects are taken from the mentioned  papers.

   The isochrones are for an age of 1 Gy and solar chemical composition; our system is probably 
older, but evolution affects only slightly the isochrone location for the mass range of interest
(the isochrone age was actually  chosen as a compromise between the age  of the
other younger systems and our object).
All the other systems but CM Dra are believed indeed to be younger: \object{CU Cnc} because of the 
association, on the basis of its space motion, with the Castor moving 
group, to which \object{YY Gem} presumably belongs, and \object{V818 Tau} as being a member of the 
\object{Hyades} cluster. 

Though the physical parameters of BW3~V38 are not accurate enough to pose constraints on
the lower main sequence calibration,  its location on the mass-radius
diagram is -- generally -- consistent with the conclusions drawn from all the other 
systems, i.e. that the theoretical models seem to underestimate the MS stellar radii
and/or overestimate the effective temperatures.  
A comparison with a larger set of models, performed by Ribas (R03) and \citet{TR02}
shows that this underestimate can be as high as 20 \% for the stellar radius, and
that most models overestimate temperatures (by ~150 K). This is regardless of the 
recent improvements in the input physics for low mass stars.
(The apparent agreement, in Fig \ref{theo}, of \object{CM Dra} parameters is obviously due to 
the difference in chemical composition of the object with respect to that of the theoretical 
isochrones; lower metallicity isochrones would correspond to smaller radii).

 The radiative properties of BW3~V38 seem in good agreement with those of the other dwarfs
in the same mass range, however,  at variance with \object{CU Cnc}, the 
temperatures could not be independently derived. 

Both BW3~V38 components are very fast rotators: the radii of Table \ref{physq} and the hypothesis
of spin--orbit synchronization yield  equatorial velocities of, respectively, 
$v_{\mathrm{1}}=131 \pm 10$ \- and $v_{\mathrm {2}}=113 \pm 15$ km~s$^{\mathrm{-1}}$.

 \citet{MMD01}  propose that  the larger radii and lower temperatures of observed 
M dwarfs  are due to the presence of  strong magnetic fields. 
There is the indication, indeed, that active dMe dwarfs (V--I $ <2.7$) are brighter 
on average by about 0.5 mag than 
dM stars of the same color \citep{ghr96}, and that the excess is even larger in
clusters: 0.7 mag in the Hyades and 1.2 mag in IC 2602. 
The non-standard stellar models of \citet{MMD01}, that include a simplified treatment of magnetic
field effects, agree with these observations, suggesting larger radii and lower temperatures
for stars hosting strong fields (up to several tens of MGauss at the base of the convective zone).
In addition in their models the limit of transition to a fully convective structure is shifted 
to masses lower than the standard model typical value ~ 0.35/0.40 M$_\odot$.
As a consequence the  $\alpha\Omega$ dynamo can still be at work for these  mass values, and
the AML through magnetic braking as well. The reason of the dearth of close systems with 
M type components shall therefore be found in the timescale of the AML  process, rather than 
in its inhibition.
 
\section{An outline of BW3~V38 orbital evolution}

 With the physical parameters of Table \ref{physq} the angular momentum content of 
BW3~V38 is $H_{\mathrm{orb}}=1.3 \cdot 10^{51}$ in cgs units (without the  spin AM that
amounts anyway to $\leq 10^{50}$ cgs, i.e. less than 10\%). This is presumably
only a fraction of the original content.

 There are good reasons to believe that systems like BW3~V38 form
with a much larger separation and evolve, later on, toward closer orbits. 
The current knowledge of close binary formation
\citep{Gh96,B00}  rules  out the fission process that does not seem able to produce  stable
binary configurations. The other ``in situ" formation process, 
capture and hardening of the binary by ejection of a component from a triple system, is rare
and inefficient in the field and works against the presence of small mass stars in the 
close pair, as  it is  generally the less massive star that is ejected.

The hydrodynamical calculations of binary formation by fragmentation of \citet{BBB02} indicate 
that proto-binaries form with large separations ($\geq$ 10 AU) and go through a phase of
accretion and orbital evolution towards tighter orbits ( $\sim$1 AU). Subsequent
less known phases of orbital evolution, presumably due to interaction with the matter 
still present in the system, should finally produce really close binaries.

On  observational grounds, \citet{Melal01} report in  their collection of  data for known  
PMS spectroscopic binaries that the shortest orbital period (relevant to a G5+K1 system)  
is $P=1\fd 67$.  

As a thorough description of the formation processes from the initial cloud to the Pre Main
Sequence (PMS)  is not yet available,  a rough estimate of a minimum separation can be 
derived by the simple argument of the orbital semi-axis necessary to accommodate the 
pre-main-sequence (PMS) progenitors of the binary. The PMS tracks of \citet{DM94} for a 
0.4 M$_\odot$ star  give a radius, during the D-burning phase, of $\sim 3$ R$_\odot$, corresponding to
a period of $\sim 1 \fd 5$.

      \begin{table}
      \caption[]{Orbital evolution parameters}
         \label{models}
     $$ 
\begin{tabular}{lllll}
\hline
\noalign{\smallskip}
Model  &   $\alpha$ & $k^2$ & $i$ & $\tau_{\mathrm{M}}$ (Gy) \\ 

\noalign{\smallskip}
\hline
\noalign{\smallskip}
A   	& 0.25   &   0.2  & 30\degr  & 2.7 \\
B		& 0.25   &   0.2  & 90\degr  & 4.5 \\  		
C		& 0.25   &   0.1  & 90\degr  & 8.9\\  		
D		& 0.00   &   0.1  & 90\degr  & 9.8 \\  		
\noalign{\smallskip}
\hline
\noalign{\smallskip}
\end{tabular}
     $$ 
   \end{table} 

Several authors have computed the orbital period evolution  of a synchronized 
binary under the influence of AML \citep{OV82,R83,MV91,CM92,Ste95,MR99} developing
an initial suggestion by \citet{VV79}.  Most works focused on deriving the slope of the 
braking law, without directly computing
the amount of AML. Moreover, the greatest part of the abovementioned works concerns  
the better known case of systems with solar-type components.
We can try at least to estimate the expected evolution of the orbital period using
the available data on M dwarfs.

The spin-down of a single late-type star can be described by:
 \begin{equation}
  \dot{\omega} = - c_{\mathrm{sd}} \: \omega^\alpha 
  \label{spindown}
 \end{equation}
where $c_{\mathrm{sd}}$ and $\alpha$ are constants and $\omega$ is the angular velocity. 
An elementary integration yields ($\alpha \neq 1$):
\begin{equation}
\omega = \omega _\mathrm{o} \left ( 1 - \frac{t-t_\mathrm{o}}{(1-\alpha)\tau _{\mathrm{sd}}} \right )^{1/(1-\alpha)}
\label{intsd}
\end{equation} 
where the quantity
\begin{equation}
\tau_{\mathrm{sd}} =\left[\frac{\omega}{\dot{\omega}} \right]_{\mathrm{t=t_o}} = \frac{\omega _\mathrm{o}^{1-\alpha}} { c_{\mathrm{sd}}}
\label{tausd}
\end{equation} 
is the spin-down timescale (for \ $t=t_o$). 
In a close binary thanks to synchronization (which can be considered as instantaneous, 
taking place on a much shorter timescale than AML, see for instance \citet{MV91}) 
the loss of spin angular momentum from both components: 
\begin{equation}
\dot{H}_{\mathrm{sp}}= k^2 (m_1 r_1^2 + m_2 r_2^2) \: \dot{\omega} = - k^2 (m_1 r_1^2 + m_2 r_2^2) \: c_{\mathrm{sd}} \: \omega^\alpha 
\label{dHsp}
\end{equation}
can be equated to the orbital angular momentum loss:
\begin{equation}
\dot{H}_{\mathrm{orb}}=-\frac{1}{3} G^{2/3} q \,(1+q)^{-2} m_\mathrm{T}^{5/3}\: \omega^{-4/3} \dot{\omega}
\label{dHorb}
\end{equation}
with  $m_\mathrm{T}$ being the total mass. 
It is assumed that $\omega _{\mathrm{orb}} \equiv \omega _{\mathrm{sp}} \equiv \omega$, and that the masses,
the radii and the gyration radius, $k^2$, do not appreciably change with time. 
Besides, the small contribution of the spin to the total AM (the quantity to which 
Equation \ref{dHorb} should apply)  is neglected. 

    \begin{figure}
	\resizebox{\hsize}{!}{\includegraphics{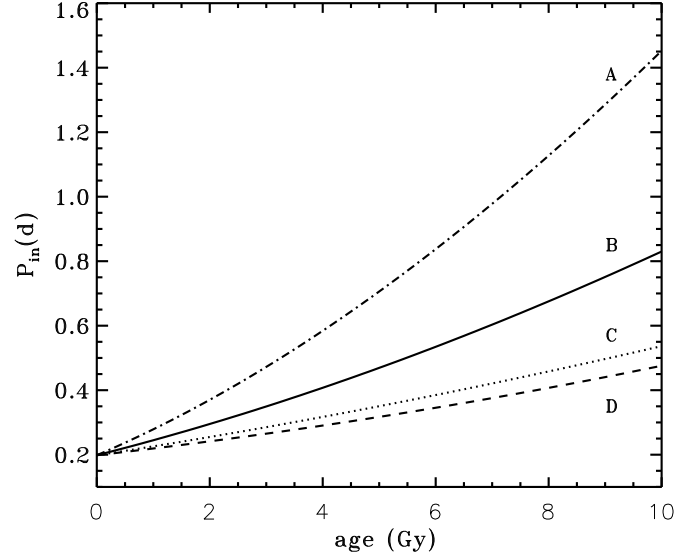}}
      \caption{The initial period, as function of system age, that is required to
	  reach by AML the present period of BW3~V38, according to the  scheme proposed
	  in the text. The curves  correspond to different values of the parameters entering
	  the computation, as indicated  in Table \ref{models}.} 
	     \label{pev}
   \end{figure}

From Eqs (\ref{dHsp}, \ref{dHorb}), and introducing the orbital period,
we get:
\begin{equation}
\dot{P}= - c \; P^{2/3-\alpha} 
\label{dpdt}
\end{equation}
with 
\begin{equation}
c= \frac{3\: (2 \pi)^{\alpha+1/3}\: k^2 \:(m_1 r_1^2 + m_2 r_2^2)\: c_{\mathrm{sd}}\: (1+q)^{2}}{G^{2/3}\: m_\mathrm{T}^{5/3}\: q}
\label{costc}
\end{equation}
 Equation (\ref{dpdt}) can be integrated once  suitable values for $\alpha$, $c_{\mathrm{sd}}$ and 
 $k^2$ are chosen. 
 
   \citet{delal98} derive that the spindown timescale of field M dwarfs at spectral 
   type M4 is 
 ``a significant fraction of  the age of the young disk ($\sim$3 Gy)", a
reasonable choice for M3 components is $\tau_{\mathrm{sd}}\simeq~1$ Gy. 

To derive  $c _{\mathrm{sd}}$ from Eq. \ref{tausd} we also need  a value for $\omega_\mathrm{o}$. The  
abovementioned paper 
gives  the observed shortest rotation period ($P/ \sin i $) of a field M4 dwarf 
belonging to the young disk population (G 165-08), its value being about 7$^\mathrm{h}$.  
 This value is derived from $v \sin i$ measurements, but it is reasonable to  assume
that the fastest observed rotation corresponds as well to
high inclination. In the following we also computed, as a check,  an extreme case with 
a very low inclination value. 

To choose $\alpha$ we recall that both the studies on
single stars and on binaries with late type components indicate   saturation of the
braking law. The saturation boundary at $v \geq 10$ km~s$^{-1}$ corresponds, 
with the radii of BW3~V38 components, to an orbital period $\sim 2\fd 6 $, meaning that
most if not all of its period evolution must have taken place in a saturated regime. 
\citet{MR99}, from a study of the period distribution of late type
binaries with equal components selected from the OGLE~I database, find a ``best value" at 
$\alpha = 0.25$. Previous studies, (as \citet{MV91} and \citet{CM92}), suggest $\alpha \simeq 0$.   
The models and the analysis of M dwarf data suggest  similar slopes, with, this in case, a shift 
of the saturation boundary to longer periods with decreasing mass \citep{Silal00,Pizzal03}.
Finally, the value of $k^2$ does not vary much with mass along the main sequence, going from
$k^2=0.1$ for a solar type star to $k^2=0.2$ for a fully convective brown dwarf \citep{AB02}. 

 Equation \ref{dpdt} was integrated with the different sets of  parameters 
of Table \ref{models} and the results are shown in Figure \ref{pev}. 
The curves  were obtained integrating the equation backwards in time and taking as a boundary
condition the present period of BW3~V38. The figure gives, therefore, as  function of 
the system age, the initial period from which our system could have evolved to the
present configuration.  The extreme case  with   $i= 30 \degr$ corresponds of course to
the fastest evolution. 
The value of $\tau_{\mathrm{sd}}$ was kept fixed, but any larger value  would make the 
evolution still slower.

In all cases period decrease is rather slow.  The evolution of a system similar to CU Cnc  (i.e., $P_{\mathrm{in}}=2 \fd 77$)
would take, to reach the BW3~V38 configuration, a time larger than  the age of the Universe 
(assumed to be 14 Gy).
 
  One may wonder if other AML mechanisms, such as AML by gravitational radiation,  could
have significantly contributed  to the orbital evolution of the system.
The  time-scale of period change due to gravitational waves can be estimated from \citep{PT72}:
\begin{equation}
\tau _{\mathrm{G}}= \left | \frac{P}{\dot{P}} \right | = 2.8 \times 10^7 
\frac{(m_{\mathrm{1}} + m_{\mathrm{2}} )^{1/3}} {
{m_{\mathrm{1}} m_{\mathrm{2}} }} P^{8/3}
\label{taug}
\end{equation}
with  $\tau _{\mathrm{G}}$ in years and $P$ in hours. For an initial period of the order 
of a day  (and  using the masses from Table \ref{physq}), we get 
$\tau _{\mathrm{G}}\simeq 7 \times 10^{11}$ y, while, for the current period value,  
$\tau _{\mathrm{G}}\simeq 9.4 $~Gy.

For comparison purposes, a timescale of AML via magnetic braking, $\tau_{\mathrm{M}}=|P/\dot{P}|$, 
can readily be derived from Eq.~\ref{dpdt} and its value,  for the current period, varies in 
the  range  2.7 -- 9.8 Gy (last column of Table \ref{models}).  The contribution of gravitational radiation
could, therefore, be relevant in the {\em future} orbital evolution of the system, and 
only in the cases where the magnetic braking has the minimum efficiency.  However - given  the 
steep dependence of $\tau _{\mathrm{G}}$ on $P$  - its contribution was negligible during  the 
past evolution of the binary.
      
  In conclusion, in the outlined scenario, if BW3~V38 has an age typical of an intermediate 
disk population (say, 5 Gy) its initial period could not have exceeded  
$P_{\mathrm{in}}\sim 1^\mathrm{d}$  and, even with all the uncertainties related to the  
AML description, it is not surprising that systems similar to it are  very rare.
  
\section{Conclusions}

The analysis of the spectroscopic data of BW3~V38, in spite of their moderate quality, 
significantly increases our knowledge of this interesting system.

The classification of the components as M3 dwarfs (which was previously
based only on the V-I color from OGLE-I) is now confirmed. Both components are indeed 
dMe stars, with strong chromospheres, as indicated by the H$\alpha$ emission profiles. 
The H$\alpha$ equivalent widths suggest  an average activity level, when compared to 
similar binaries. This is an indication that the system is older than similar systems.
Thanks to BW3~V38 we have  clear evidence that binaries with M dwarfs and very short 
periods are rare objects but do exist. If they form by AML from wider orbits, the AML 
timescale will then be  $\leq$ the age of the Galaxy. 

The simultaneous radial and light curve solution provided a consistent set of 
absolute elements. The components are very similar stars, with the slightly more massive
one  very close to the limiting Roche surface. 
The system can be considered  a tighter twin of \object{CU Cnc}.
 
 The derived masses and radii agree - within the limits of the large uncertainty - with
the relations obtained from the 
M-type components of the other known systems, which indicate a discrepancy, of the  
order of 10\%, between the theoretical models and the values determined from eclipsing
binary stars. It has been suggested that the effect could be related to the presence
of strong magnetic fields, as dMe stars are indeed brighter than the inactive dM dwarfs.
In addition  the BW3~V38 components are very fast rotators (~130 km~s$^{-1}$) and the 
larger radii could also be related to this property.
 
A simple estimate of the orbital evolution of the system, taking into account the 
latest findings on the spin down of late type dwarfs, shows that
the AML loss process is rather slow (the evolution from a configuration similar
to that of \object{CU Cnc} would take, in the fastest case, $\sim$ 14 Gy). In this
frame the rareness of binaries similar to BW3~V38 can easily work. 

It is clear that the system deserves better quality observations. It could be
a fundamental system to better constrain the lower MS calibration, and to study
the effects of magnetic fields and fast rotation, but higher S/N
spectra are needed.
These will require a  telescope of the 8m class. We hope that the results 
obtained with this study will trigger the interest in this unique object, 
and  qualify it as a priority target for the immediate future.

%__________________________________________________________________

\begin{acknowledgements}
The collection of data for this system was particularly difficult, as  weather
and technical problems impaired several efforts in this direction. CM  especially
thanks Slavek Rucinski, for continuous encouragement and for  his help during a 
first  attempt at CFHT, that unfortunately did not produce any useful photons. 
The team of ESO-NTT is also acknowledged for precious advice during and after the 
two observing runs at ESO. 
 The authors are also indebted  to Vincenzo Testa and Gian Luca Israel
for discussions on data reduction techniques, to Francesca D'Antona for providing 
new isochrones and to the anonymous referee for useful comments.
 This research project was  funded by MIUR/Cofin and F-INAF programs and by  
ESA-Prodex under contract 15448/01/NL/SFe(IC)-C90135.
\end{acknowledgements}

\bibliographystyle{aa}

\end{document}